\documentclass[prd,12pt,nofootinbib,superscriptaddress]{revtex4-2}
\usepackage{bm}
\usepackage{amsmath}
\usepackage{amssymb}
\usepackage{graphicx}
\usepackage{float}
\usepackage{subfigure}
\usepackage{hyperref}
\usepackage{CJK}
\hypersetup{
	colorlinks=true,
	linkcolor=red,
	citecolor=blue,
}

\def\lsim{\mathrel{\rlap{\lower4pt\hbox{\hskip0.5pt$\sim$}}
    \raise1pt\hbox{$<$}}}         
\def\gsim{\mathrel{\rlap{\lower4pt\hbox{\hskip0.5pt$\sim$}}
    \raise1pt\hbox{$>$}}}         

\begin{document}
\title{Detecting new fundamental fields with Pulsar Timing Arrays}
\author{Chao Zhang}
\email{zhangchao666@sjtu.edu.cn}
\affiliation{School of Aeronautics and Astronautics, Shanghai Jiao Tong
University, Shanghai 200240, China}

\author{Ning Dai}
\email{Corresponding author. daining@hust.edu.cn}
\affiliation{School of Physics, Huazhong University of Science and Technology, Wuhan, Hubei
430074, China}

\author{Qing Gao}
\email{Corresponding author. gaoqing1024@swu.edu.cn}
\affiliation{School of Physical Science and Technology, Southwest University, Chongqing 400715, China}

\author{Yungui Gong}
\email{yggong@hust.edu.cn}
\affiliation{School of Physics, Huazhong University of Science and Technology, Wuhan, Hubei
430074, China}
\affiliation{Department of Physics, School of Physical Science and Technology, Ningbo University, Ningbo, Zhejiang 315211, China}

\author{Tong Jiang}
\email{jiangtong@hust.edu.cn}
\affiliation{School of Physics, Huazhong University of Science and Technology, Wuhan, Hubei
430074, China}

\author{Xuchen Lu}
\email{Luxc@hust.edu.cn}
\affiliation{School of Physics, Huazhong University of Science and Technology, Wuhan, Hubei
430074, China}

\begin{abstract}
Strong evidence of the existence of the Stochastic Gravitational-Wave Background (SGWB) has been reported by the NANOGrav, PPTA, EPTA and CPTA collaborations.
The Bayesian posteriors of the Gravitational-Wave Background (GWB) amplitude and spectrum are compatible with current astrophysical predictions for the GWB from the population of supermassive black hole binaries (SMBHBs).
In this paper, we discuss the corrections arising from the extra scalar or vector radiation to the characteristic dimensionless strain in PTA experiments and explore the possibility to detect charges surrounding massive black holes, which could give rise to SGWB with vector or scalar polarizations.
The parametrized frequency-dependent characteristic dimensionless strain is used to take a Bayesian analysis and  the Bayes factor is also computed for charged and neutral SMBHBs.
The Bayesian posterior of GWB tensor amplitude is $\log_{10} A_T=-14.85^{+0.26}_{-0.38}$ and spectral exponent $\alpha=-0.60^{+0.32}_{-0.36}$.
The Bayesian posterior for vector or scalar amplitude $A_{V, S}$ is nearly flat and there is nearly no constraint from the current observation data.
The Bayesian factor is $0.71$ far less than 100, so the current observation can not support the existence of the charged SMBHB.
\end{abstract}

\maketitle
\section{Introduction}
The first direct detection of gravitational wave (GW) event GW150914 provides a new perspective on understanding gravity in nonlinear and strong field regimes, marking the inception of GW astronomy \cite{Abbott:2016blz, TheLIGOScientific:2016agk}.
To date, over 90 GW events resulting from binary star mergers have been detected \cite{LIGOScientific:2018mvr, LIGOScientific:2020ibl, LIGOScientific:2021usb, LIGOScientific:2021djp}.
In addition to these individual and instantaneous GW sources, 
there is a continuous interest in the stochastic gravitational-wave background (SGWB), whose signals are from multiple continuous GW sources.
The recent data is released by multiple Pulsar Timing Array (PTA) experiments, including the North American Nanohertz Observatory for Gravitational Waves (NANOGrav) \cite{NANOGrav:2023gor, NANOGrav:2023hde}, the European PTA (EPTA) \cite{Antoniadis:2023ott, Antoniadis:2023lym, Antoniadis:2023xlr}, the Parkes PTA (PPTA) \cite{Reardon:2023gzh, Zic:2023gta, Reardon:2023zen}, and the Chinese PTA (CPTA) \cite{Xu:2023wog}, show evidence for Hellings-Downs angular correlations \cite{Hellings:1983fr}, which indicating the observed stochastic common spectrum can be interpreted as an SGWB.
The observations strongly support the hypothesis that the signals, with a frequency spectrum emitted by supermassive black hole binaries (SMBHBs).

There are supermassive black holes (SMBHs) with masses ranging from $10^5$ to $10^{10}~M_{\odot}$ located at the center of most the galaxies \cite{Kormendy:1995er}. 
These SMBHs can form SMBHBs after their host galaxies merged with other galaxies.
The GWs emitted by the SMBHBs, contribute to a noise-like broadband signal in the nHz range \cite{DeRocco:2023qae, Ghoshal:2023fhh}. 
Due to the gravitational reaction, the GWs frequencies evolve slowly, and
the frequency spectrum follows a power-law relationship, with $h_c(f) \propto f^{-2/3}$ at $2\sigma$ by the latest data \cite{NANOGrav:2023hfp, NANOGrav:2023hvm}.
However, environmental and statistical effects may lead to different predictions \cite{Sesana:2008mz, Kelley:2016gse, Ellis:2023owy,kocsis2011gas}.
The expected amplitude of the astrophysical background is subject to an order of magnitude uncertainty due to several factors, 
such as the distribution of masses of the SMBHBs, the eccentricity of the binary-orbits, and the redshift. 
Moreover, the possible contribution of modified gravity theories and the charges and spin of the SMBHs may adds the uncertainty in the amplitude estimation \cite{Yunes:2013dva}.
There are also numerous interpretations based on cosmological sources, including cosmic strings and domain walls \cite{An:2023idh,Qiu:2023wbs,Zeng:2023jut}, first-order phase transitions \cite{NANOGrav:2021flc,Gouttenoire:2023naa}, and primordial fluctuations \cite{Dandoy:2023jot,Zhao:2023xnh,Ferrante:2023bgz,Cai:2023uhc,Inomata:2023zup,Cai:2023dls}.
The SGWBs generated by cosmological sources are significant for understanding new physics beyond the Standard Model and providing insights into the primordial universe.

The observation of SGWB is also important for the study of new fields such as scalar and vector fields around SMBHs.
The BHs can be charged with scalar charges in scalar-tensor theories \cite{Campbell:1991kz, Mignemi:1992nt,Kanti:1995vq,Yunes:2011we,Kleihaus:2011tg,Sotiriou:2013qea,Sotiriou:2014pfa,Antoniou:2017acq,Doneva:2017bvd,Silva:2017uqg,Cardoso:2020iji}, or the vector charges of some dark matter models with multi-charged components \cite{Holdom:1985ag, Cardoso:2020iji, Cardoso:2016olt}.
Even the BHs may carry electromagnetic charges, as predicted by the model of Kerr-Newman BHs \cite{Newman:1965my}.
It's shown that the future GW observations of space and ground detectors can impose severe limits on the charge of BH \cite{Scharre:2001hn, Barausse:2016eii, Maselli:2021men, Maselli:2020zgv, Liang:2022gdk, Zhang:2023vok, Zhang:2022rfr}.
The discovery of nHz SGWB opens a new window for testing astrophysical processes and whether extra polarization emission exists.
In addition to tensor GW emission, binaries may also emit scalar and vector radiation. 
These additional forms of emission can occur when the BHs comprising the binaries possess scalar charges or vector charges.
The extra polarizations such as vector and scalar polarization would give rise to SGWB.
In this study, we focus on the possibility of detecting the extra polarization power spectrum from charged SMBHBs within the new results of NANOGrav and PPTA.

The paper is organized as follows: in Sec. \ref{s-v radiation} we discuss the corrections arising from the extra scalar or vector radiation to the characteristic dimensionless strain in PTA experiments.
The parameterized frequency-dependent characteristic dimensionless strain is given based on the important approximation to the overlap reduction function valid within the frequency range of current PTA experiments.
In Sec. \ref{data-a}, we perform the Bayesian analysis to estimate the signals and give the posterior probability distribution of GWB amplitude and spectra from SMBHs with and without charge.
Conclusions are given in section~\ref{concl}.

\section{SGWB in the presence of scalar or vector radiation}
\label{s-v radiation}
In the theory of massless fields including scalar charge or vector charge \cite{Cardoso:2020iji}, we consider SMBHB components carrying some additional charge.
The scalar-tensor theories of gravity include scalar charge through spontaneous scalarization \cite{Damour:1993hw, Silva:2017uqg, Doneva:2017bvd}, while some dark matter models with millicharged components include vector charge (electromagnetic charge) \cite{Cardoso:2016olt}.
The Lagrangian density of gravity for binary components is given by
\begin{eqnarray}
\mathcal{S}&=&\int d^4x \frac{\sqrt{-g}}{16\pi G}\bigg[R-\frac{1}{2}g^{\mu \nu}\Phi_{,\mu} \Phi_ {,\nu}-\frac{1}{4}F^{\mu\nu}F_{\mu\nu}\nonumber\\
&-&\frac{1}{\sqrt{-g}}\sum_{j=1}^{2}(m_j+ 4 \pi m_j q^{0}_j \Phi) \int d\lambda \sqrt{-g_{\mu \nu}\dot{z_j}^\mu \dot{z_j}^\nu}\delta^4(x-z_j)\nonumber\\
&-&\frac{4\pi}{\sqrt{-g}}\sum_{j=1}^2 m_j q^{1}_j A_\alpha \int d\lambda\, \dot{z}_j^\alpha\delta^4(x-z_j)\bigg]\,,\label{Action_total}
\end{eqnarray}
where $\Phi$ is a massless scalar field, $A_\mu$ is a massless vector field, $F_{\mu\nu}=\nabla_{\mu}A_{\nu}-\nabla_{\nu}A_{\mu}$ is the field strength, $m_1$ and $m_2$ are the masses of two objects of the compact binary.
The parameter $q^{s}_i$ is the charge in unit mass carried by each binary component where $s=0$ corresponds scalar field and $s=1$ corresponds vector field.

Within the bandwidth of PTA, the background is mainly from the inspiral stage.
The power of tensor, vector, and scalar emission from inspiraling charged binary systems has been studied in \cite{Cardoso:2020iji, Du:2018txo}.
Contrary to the tensor case, the vector and scalar fields contribute monopole and dipole radiation as well as quadrupole radiation.
In the limit of vanishing eccentricity, the scalar and vector energy spectrum from the monopole radiation and quadrupole can be negligible. 
Thus the main contribution of the scalar and vector field to the background is from dipole radiation
\begin{equation}
\left\langle \frac{dE_{V,S}}{dt} \right\rangle=-\frac{4(s+1)(q_1^s-q_2^s)^2}{3}\frac{G^2m_1^2m_2^2}{a^4}.
\end{equation}
 For the tensor field, the main contribution to the background is from quadrupole radiation
\begin{equation}
\left\langle \frac{d E_T}{dt} \right\rangle=-\frac{32}{5}\frac{G^4m_1^2m_2^2M}{a^5},
\end{equation}
where  $M=m_1+m_2$ is the total mass and $a$ is the orbital semimajor axis.
For binaries dominated by gravitational interaction, the orbital frequency $F$ satisfies Kepler's law 
\begin{equation}
F=\sqrt{\frac{GM}{a^3}}.
\end{equation}
In the limit of vanishing eccentricity, the orbital frequency $F$ and the GW frequency $f$ is related by $f=j F$ for $j=1,2$.
The rate of change of the orbital frequency due to GW emission is 
\begin{equation}
    \dot{F}=\frac{48\pi^{8/3}G^{5/3}}{5}\frac{m_1m_2}{M^{1/3}}(2F)^{11/3}.
\end{equation}
The energy spectrum is derived from the relation to the power,
\begin{equation}
\frac{d E_I}{df}= \frac{1}{\dot{f}}\left\langle \frac{d E_I}{dt} \right\rangle,
\end{equation}
where the modes of type $I=T, V, S$ represents tensor, vector or scalar modes respectively. For the tensor part from quadrupole radiation $(j=2)$, the energy spectrum is 
\begin{equation}
\frac{d E_T}{df}=\frac{G^{2/3}\pi^{2/3}m_1m_2}{3M^{1/3}}f^{-1/3}.    
\end{equation}
For the vector and scalar part from dipole radiation $(j=1)$, the energy spectrum is
\begin{equation}
\frac{d E_{V,S}}{df}=\frac{5(s+1)(q_1^s-q_2^s)^2}{72}\frac{m_1m_2}{M}f^{-1}. 
\end{equation}

We expect the presence of the tensor, vector, and scalar GW would give rise to a stochastic background with tensor, vector, and scalar polarizations, which is described by the dimensionless energy density spectrum:
\begin{equation}
    \begin{split}
\Omega_T(f)&=\frac{1}{\rho_c}\frac{d\rho_T}{d\ln f},\\
  \Omega_V(f)&=\frac{1}{\rho_c}\frac{d\rho_V}{d\ln f},\\
\Omega_S(f)&=\frac{1}{\rho_c}\frac{d\rho_S}{d\ln f}, 
\end{split}
\end{equation}
where $\rho_c=3H_0^2/8\pi G$ is the critical density and $H_0$ represents today's Hubble expansion parameter. The quantities $\rho_T$,  $\rho_V$, and $\rho_S$ relate to tensor, vector, and scalar energy density in the frequency, respectively.
The energy density spectrum of the produced SGWB can be obtained from the emission spectrum of a single SMBHB merger event \cite{Phinney:2001di}
\begin{equation}
\Omega_I(f)=\frac{1}{\rho_c}\frac{d\rho_I}{d\ln f}=\frac{f}{\rho_c}\int_0^{z_{\rm max}}dz \frac{R_m(z)}{(1+z)H(z)}\frac{d E_I}{df}(f_z),
\end{equation}
where $f_z=(1+z)f$ is the frequency at emission, $R_m(z)$ is the SMBHB merger rate per comoving volume at redshift $z$ \cite{Klein:2015hvg} and $z_{\rm max}=10$ is the redshift cutoff.
We adopt the $\Lambda \rm CDM$ cosmological model with 
\begin{equation}
    H(z)=H_0\left[\Omega_{m0}(1+z)^3+(1-\Omega_{m 0})\right],
\end{equation}
where the cosmological parameters are chosen as the Planck 2018 results: $H_{0} = 67.27$ km/s/Mpc, and $\Omega_{m0}=0.3166$ \cite{Planck:2018vyg}.
It is customary to use the GW strain power spectrum as a function of characteristic frequency $h_{c,I}(f)$ for PTA searches.
The characteristic dimensionless strain $h_{c,I}(f)$ is related to $\Omega_I(f)$ by
\begin{equation}
\label{hcI}
\Omega_I(f)=\frac{2\pi^2}{3H_0^2}f^2h_{c,I}^2(f).
\end{equation}
The GW strain power spectrum is typically approximated as the power-law form at a reference frequency $f_{\rm yr}=1 \rm{yr}^{-1}$, with amplitude and spectral index given by $A_{c,I}$ and $\alpha_{c,I}$ respectively:
\begin{equation}
 h_{c,I}(f)=A_{c,I}\left(\frac{f}{f_{\rm yr}}\right)^{\alpha_{c,I}},  
\end{equation}
For the tensor part, we have $\alpha_{c,T}=-2/3$,
and $\alpha_{c,V}=\alpha_{c,S}=-1$ for the vector or scalar part.
The observable data for PTA data is the two-point correlation function of quantity $z$.
The full two-point function from all polarization contributions takes the schematic form
\begin{equation}
\langle \hat{z}(f) \hat{z}^*(f)\rangle  =\Omega_T \Gamma_T+  \Omega_V \Gamma_V+  \Omega_S\Gamma_S,
\end{equation}
where $\Gamma_T$, $\Gamma_V$, and $\Gamma_S$ represent the overlap reduction function for tensor, vector, and scalar polarization \cite{Liang:2021bct}.
It is convenient to define an "effective" energy density to describe the effects caused by vector polarization and scalar polarization
\begin{equation}
\Omega_{\rm eff}=\Omega_T+\frac{\Gamma_{V}}{\Gamma_T}\Omega_{V,S}+\frac{\Gamma_S}{\Gamma_T}\Omega_S.
\end{equation}
Since $\Gamma_T$, $\Gamma_V$, and $\Gamma_S$ are independent of frequency in the bandwidth of PTA, we can parametrize the frequency-dependent characteristic dimensionless strain $h_{c}$ in the form,
\begin{equation}
h_{c}=A_T\left(\frac{f}{f_{\rm yr}}\right)^{\alpha}+A_{V,S} \left(\frac{f}{f_{\rm yr}}\right)^{-1}.
\end{equation}
In this situation, a population of GW-driven circular SMBHBs produces a spectrum with $\alpha=-2/3$ and amplitude $A_T \approx 10^{-15}$ \cite{Antoniadis:2023xlr}.
The amplitude $A_{V, S}$ is dependent on the overlap reduction function, charges carried by the binary component, and SMBHB population model.

\section{Data Analysis and Results}
\label{data-a}
\begin{figure}
\centering
\includegraphics[width=0.9\linewidth]{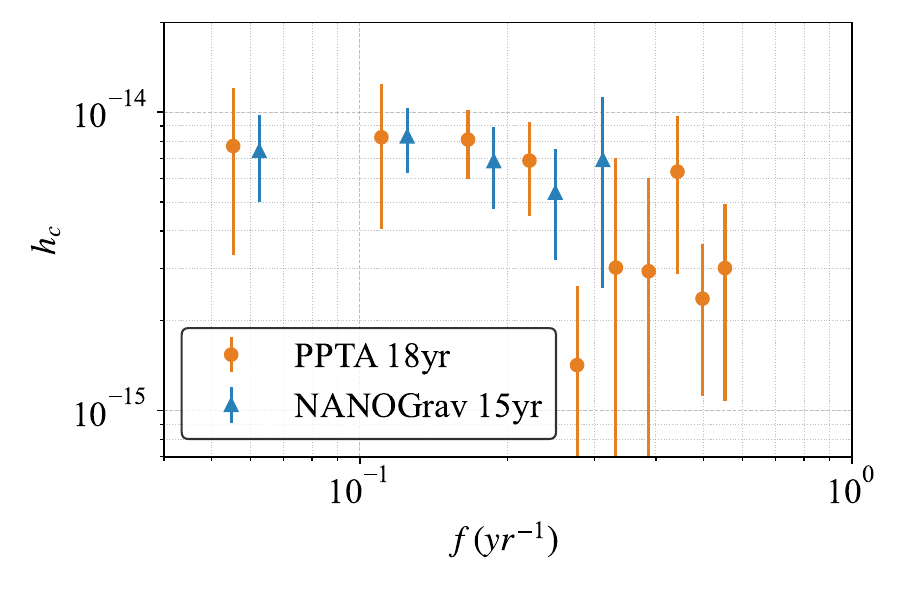}
\caption{Characteristic strain signals from SMBHBs are displayed. The blue points and orange points are the results that come from NANOGrav 15yr~\cite{NANOGrav:2023gor} and PPTA 18yr~\cite{Reardon:2023gzh} observations.}
\label{hc}
\end{figure}

\begin{figure}
\centering
\includegraphics[width=0.8\linewidth]{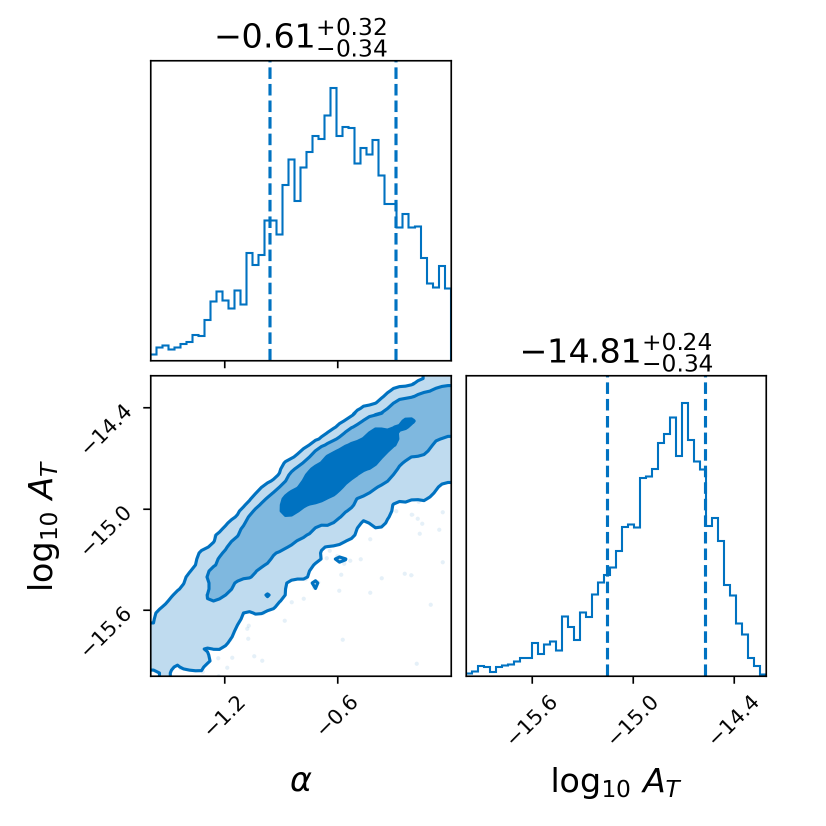}
\caption{The probability regions for the amplitude and slope of a power-law fit to the observed SGWB signal.}
\label{hdhc}
\end{figure}

\begin{figure}
\centering
\includegraphics[width=0.8\linewidth]{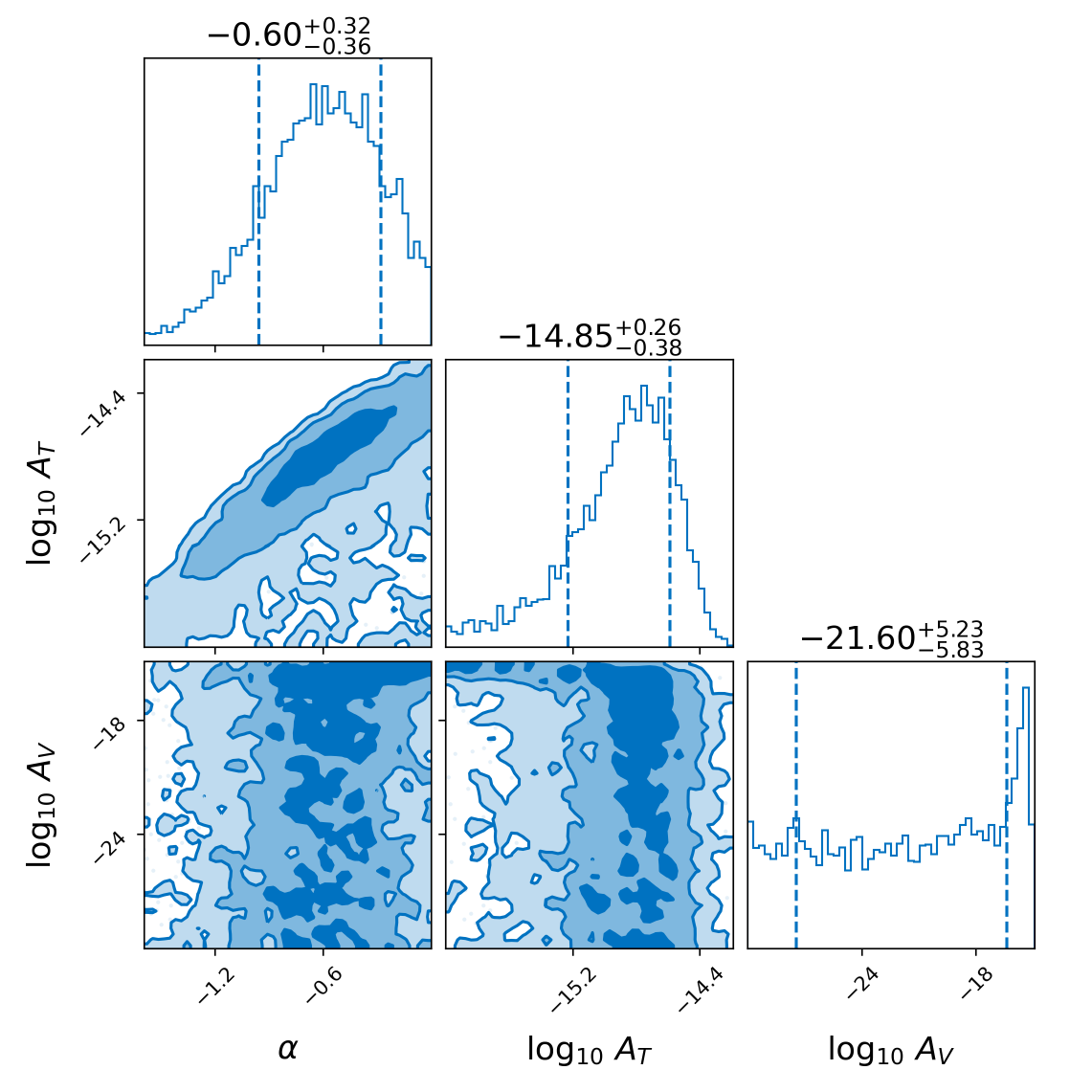}
\caption{The probability regions for the amplitudes $A_T$, $A_V$, and slope $\alpha$ of a power-law fit to the observed SGWB signal with vector polarization.}
\label{CBH}
\end{figure}

Recently, the NANOGrav collaboration, PPTA collaboration, EPTA collaboration, and CPTA collaboration have published their measurements on SGWB.
The results show the presence of SGBW with a power law spectrum is favored over a model with only independent pulsar noises.
Following the paper \cite{Shen:2023pan}, the first ten and first five  excess timing delays measured by PPTA collaboration \cite{Reardon:2023gzh} and NANOGrav collaboration \cite{NANOGrav:2023gor} are converted to the characteristic strain~\cite{Ratzinger:2020koh}
\begin{equation}
    {\rm residual}(f) = \frac{1}{4\pi^2 f_{\rm yr}} \left( \frac{f}{f_{\rm yr}}\right)^{-3/2} h_{\rm c}(f).
\end{equation}
Figure \ref{hc} gives the median values and errors of the observed data from the NANOGrav collaboration and PPTA collaboration.
We use the maximum likelihood to explore the implications and constraints of the extra polarization background caused by the existence of charges on SMBHBs.
The likelihood $\mathcal{L} = p\left(h_{{\rm c},i}|\Theta\right)$ with two data sets is defined
\begin{equation}
\label{deflnL}
    \ln {\mathcal L}(\Theta) =-\frac{1}{2} \sum_{\rm \{NANOGrav, PPTA\}} \left[ \frac{h_{{\rm c},i} - h_{\rm c}(f_i;\Theta)}{\sigma_{i}} \right]^2,
\end{equation}
where $h_{{\rm c},i}$ and $\sigma_{i}$ are the median values and errors of the observed data, $h_{\rm c}(f_i;\Theta)$ is the modeled strains at frequency $f_i$ with parameters $\Theta=(A_T, \alpha, A_{V,S})$.
Because of the same formula for vector and scalar, we can combine them and analyze only one situation.
The posterior distribution for the parameters $\Theta$ is
\begin{equation}
\label{defBayes}
p\left( \Theta | h_{{\rm c},i} \right) = \frac{p(h_{{\rm c},i} |\Theta)p(\Theta)}{p(h_{{\rm c},i})},
\end{equation}
where $p(\Theta)$ is the prior on the parameters and $p(h_{{\rm c},i})$ is the evidence.
We use the public code Bilby \cite{Ashton:2018jfp}
to perform Bayesian analyses with Eqs. \eqref{deflnL} and ~\eqref{defBayes}.
We choose the sampler Dynesty \cite{10.1214/06-BA127} and 1000 live points for nested sampling and we obtain the posteriors of the physical parameter $\Theta$.
We first fit the SMBHB model without charge.
For our fiducial power-law model and a log-uniform amplitude prior, the Bayesian posterior of SGWB amplitude at the customary reference frequency $1\rm{yr}^{-1}$ is $\log_{10} A_T=-14.81^{+0.24}_{-0.34}$, which is compatible with current astrophysical estimates for the SGWB from SMBHBs.
Also, $\alpha=-0.61^{+0.32}_{-0.34}$ is compatible with current astrophysical estimates $-2/3$ for the SGWB from SMBHBs.
The posterior probability distribution of SGWB amplitude $A_T$ and spectral exponent $\alpha$ are shown in Fig.~\ref{hdhc}.
Then the model for SMBHB with charges is optimized.
The posterior of SGWB amplitude and spectral exponent for three degrees of freedom including charges is nearly the same as the result from the model without charges.
The posterior probability distribution of SGWB tensor amplitude $A_T$, vector amplitude $A_V$, and spectral exponent $\alpha$ are shown in Fig.~\ref{CBH}.
The Bayesian posterior of SGWB amplitude is $\log_{10} A_T=-14.85^{+0.26}_{-0.38}$ and spectral exponent $\alpha=-0.60^{+0.32}_{-0.36}$.
As for the amplitude of vector polarization caused by charges,  the Bayesian posterior for $A_V$ is nearly flat and there is nearly no constraint from the current observation data.
To quantify whether we can distinguish charged SMBH from neutral SMBH, we take a Bayesian approach by computing the Bayes factor for the charged and neutral SMBH.
The ratio of the evidence for the signal under each model,
\begin{equation}
{\rm {BF}}(d)=\frac{p(h_{{\rm c},i}|TV)}{p(h_{{\rm c},i}|T)},
\end{equation}
where the evidence for a model with parameters $\Theta$ is 
\begin{equation}
p(d)=\int d \Theta~p_{\rm max}(h_{{\rm c},i}|\Theta)p(\Theta),
\end{equation}
where $p_{\rm max}(d|\Theta)$ is the maximized likelihood and $p(\Theta)$ is the prior.
The Bayesian factor is $\rm{BF}=0.71$.
A signal for which the Bayes factor exceeds $100$ can be understood as decisively favoring chared SMBHs than neutral SMBHs.
So the current observation data can not support the existence of the charge on SMBH.

\section{Conclusion}
\label{concl}
Due to the presence of additional charge carried by binaries in the astrophysical environment, the radiated GWs of them deviate from the predictions of general relativity.
In this paper, we estimate the probability of constraint on the charges in the astrophysical environment around the SMBHs by PTA observations.
We take the Bayesian analysis for neutral and charged SMBHB models. 
For the model without charges, based on the fiducial power-law model and the log-uniform amplitude prior, the Bayesian posteriors of SGWB are $\log_{10}A_T=-14.81^{+0.24}_{-0.34}$ and $\alpha=-0.61^{+0.32}_{-0.34}$, which are compatible with current astrophysical estimations for the SGWB from SMBHBs.
For models with charges, the results of the tensor part are almost consistent with the models without charges by $\log_{10}A_T=-14.85^{+0.26}_{-0.38}$ and $\alpha=-0.60^{+0.32}_{-0.36}$.
For the amplitude of extra polarizations caused by the charges, the Bayesian posterior for $A_V$ is nearly flat and there is nearly no constraint from the current observation data.
The Bayesian factor between the charged SMBHs model and the neutral SMBHs model is only 0.71, which is far less than 100.
Thus, the current observation can not support the existence of the charged SMBH.

\begin{acknowledgments}
We thank Dicong Liang for the useful discussions.
This work makes use of the Bilby package.

\end{acknowledgments}

%

\end{document}